\RequirePackage{fix-cm}

\documentclass[twocolumn,epjc3]{svjour3}  

\smartqed

\RequirePackage{graphicx}
\RequirePackage[colorlinks,citecolor=blue,urlcolor=blue,linkcolor=blue]{hyperref}

\usepackage{xcolor}
\usepackage{color}
\usepackage{bm}
\usepackage{amsmath}
\usepackage{cuted}
\usepackage{mathtools}
\usepackage[separate-uncertainty = true]{siunitx}

\journalname{Eur. Phys. J. C}

\begin{document}

\title{First-principle calculation of birefringence effects for in-ice radio detection of neutrinos}

\author{Nils Heyer\thanksref{upp,e1}         
        \and
        Christian Glaser\thanksref{upp,e2} 
}
\thankstext{e1}{nils.heyer@physics.uu.se}
\thankstext{e2}{christian.glaser@physics.uu.se}

\institute{Uppsala University Department of Physics and Astronomy, Uppsala SE-75237, Sweden\label{upp}}

\date{}

\maketitle

\begin{abstract}
The detection of high-energy neutrinos in the EeV range requires new detection techniques to cope with the small expected flux. The radio detection method, utilizing Askaryan emission, can be used to detect these neutrinos in polar ice. The propagation of the radio pulses has to be modeled carefully to reconstruct the energy, direction, and flavor of the neutrino from the detected radio flashes. Here, we study the effect of birefringence in ice, which splits up the radio pulse into two orthogonal polarization components with slightly different propagation speeds. This provides useful signatures to determine the neutrino energy and is potentially important to determine the neutrino direction to degree precision. We calculated the effect of birefringence from first principles where the only free parameter is the dielectric tensor as a function of position. Our code, for the first time, can propagate full RF waveforms, taking interference due to changing polarization eigenvectors during propagation into account. The model is available open-source through the NuRadioMC framework. We compare our results to in-situ calibration data from the ARA and ARIANNA experiments and find good agreement for the available time delay measurements, improving the predictions significantly compared to previous studies. Finally, the implications and opportunities for neutrino detection are discussed.
\end{abstract}

\section{Introduction}

Neutrinos are perfect cosmic messengers \cite{messenger}. Because of the ghostly nature of these peculiar elementary particles that allows them to pass through matter almost unhindered, neutrinos will provide insights into the inner processes of the most violent phenomena in our universe \cite{Astro2020NeutrinoAstronomy,sources}. Detection of neutrinos at ultra-high energies (UHE, $>$\SI{e17}{eV}) would be one of the most important discoveries in astroparticle physics in the 21st century \cite{Ackermann2022}.

However, the low expected flux of UHE neutrinos and their ghostly nature make their detection challenging. IceCube, the world’s largest neutrino telescope, has to date detected neutrinos with energies up to \SI{e16}{eV}. Though no neutrino above \SI{e17}{eV} has been detected so far, limits could be set where the strongest limits come from the IceCube \cite{IceCube_limit} and Auger \cite{Auger_limit} experiments. Current detector technologies like those of IceCube become cost-prohibitive for higher energies. Therefore, a new detection technique has been developed over the last decade where an array of radio antennas installed in the polar ice sheet searches for radio flashes generated by neutrinos interacting in the ice \cite{KRAVCHENKO2003195,ARIANNALimit2020,ARALimit2020}. The radio technology allows cost-efficient instrumentation for the monitoring of large volumes.

When high-energy neutrinos interact in ice they create a particle shower and the secondary particles generate a short radio flash via the Askaryan effect \cite{Askar,ZHS}. Using ice as the detector medium has the advantage that it is readily available in polar regions and that the attenuation length of radio signals often exceeds \SI{1}{km} \cite{attenuation}. This allows instrumenting of large volumes with a sparse array of radio detector stations. However, it also requires a good understanding of the kilometer-long propagation of radio signals through the ice to recover the neutrino properties from the observable radio flashes. Many experiments are dedicated to building such a detector and pushing for a measurement of the neutrino flux up into the EeV region. The detector technology has been successfully explored in the pilot arrays ARIANNA \cite{ARIANNALimit2020} and ARA \cite{ARA_limit}. The Radio Neutrino Observatory in Greenland (RNO-G) is the first detector of sufficient size to potentially detect the first UHE neutrino and is currently being constructed in Greenland (2021-2024) \cite{RNOGWhitePaper2021}. At the same time, an order of magnitude larger radio array is being planned as part of the IceCube-Gen2 efforts to build the next-generation neutrino observatory at the South Pole \cite{Gen2WhitePaper,GEN2_limit} with hundreds of autonomous detector stations. 

To reconstruct the neutrino properties from the detected signals, the propagation through the ice from the neutrino interaction to the antenna has to be understood to good precision. This in turn calls for a precise model of the medium in which the radio pulses propagate and its effect on the propagation. Ice has a non-uniform structure depending on, e.g., crystal fabric, density, pressure due to ice flow, or impurities like air bubbles or ash layers. One effect is biaxial birefringence which alters the propagation speed depending on the signal polarization of the radio pulses. This paper introduces a model that simulates the effect of birefringence for in-ice radio propagation. It makes detailed predictions about the pulse shapes, polarization, and arrival time for arbitrary geometries possible. The strength of the model is the mathematical foundation on which it is based. The calculation of the effective refractive indices as well as the steps of the numerical pulse propagation are well-founded in classical electrodynamics. The only free parameters in the calculation are the ice properties, i.e., the index-of-refraction including the polarization-dependent asymmetries from birefringence as a function of position. We combine the calculations with a numerical propagation code that allows the propagation of arbitrary pulse forms through the ice. This allows detailed modeling of in-situ measurements, as well as neutrino-induced radio pulses.

In this work, we use the ice-fabric measurement from the SPICE core project at the South Pole \cite{voigt} from which the dielectric tensor was derived \cite{Jordan}. We build upon previous work that studied birefringence effects on in-ice propagation \cite{Jordan,Amy} but improve it significantly. The previous models were restricted either to special geometries where the effective refractive indices could be approximated or to continuous waveforms of fixed frequency instead of short broadband pulses that are relevant for radio neutrino detection. Our model calculates an analytical solution directly derived from Maxwell's equations for arbitrary geometries and allows for the propagation of realistic pulse forms. 

We integrated the birefringence model into the NuRadioMC framework. NuRadioMC is an open-source python-based Monte Carlo code to create precise simulations of the neutrino interaction, the Askaryan emission, the radio propagation, and the detector response \cite{NuRadioMC,NuRadioReco}. The birefringence model is an extension of the ray-tracing class of the propagation simulation. The integration into NuRadioMC makes our model available and directly usable to the in-ice radio community and will allow studying the impact of birefringence on neutrino detection in future work.

The paper is structured as follows: We first present the calculation of birefringence effects from first principles and describe the numerical procedure we developed to propagate arbitrary waveforms through the ice. Then, we make predictions and compare them to existing in-situ measurements at the South Pole where we find that the ability of our model to propagate arbitrary waveforms is crucial for the interpretation of the data. Finally, we study the impact of birefringence on the radio detection of ultra-high-energy (UHE) neutrinos.

\section{Birefringence Model}
\label{sec:birefringence_model}

In this section, we derive the calculation of birefringence from first principles and describe how we integrate it into the NuRadioMC code. The only free parameters of the model are the position-dependent dielectric properties of the medium. Throughout this work, we chose a coordinate system that aligns with the symmetry of the dielectric tensor to simplify the calculation: The z-axis points in the vertical direction, the x-axis runs parallel to the direction of the horizontal ice flow and the y-direction runs perpendicular to the direction of the horizontal ice flow.

\subsection{Ice Model}
\label{section: Ice Model}

The largest influence on the index-of-refraction of ice is the ice density. Over the upper $\mathcal{O}$(\SI{200}{m}), often referred to as the firn, the density gradually changes from fluffy snow to solid ice which leads to a change of the index-of-refraction from approx. $n=1.35$ at the surface to $n=1.78$ at deeper depths. The density profile has been measured at several places around the South Pole (see Ref.~\cite{southpole_2015} for a compilation of available measurements) and the resulting index-of-refraction profiles $\langle n(z)\rangle$ can be described via an exponential function of the form

\begin{equation} 
\label{eq:density}
\langle n(z) \rangle = 1.78 - \Delta n \cdot \exp\left( \frac{z}{z_0} \right) \, ,
\end{equation}
where $\Delta n$ and $z_0$ are free parameters that are determined from density measurements of ice cores \cite{southpole_2015}, or directly from propagation times of radio waves in ice \cite{Beise2022}. 

\begin{figure}[tbp]
  \centering
  \includegraphics[width = 0.48\textwidth]{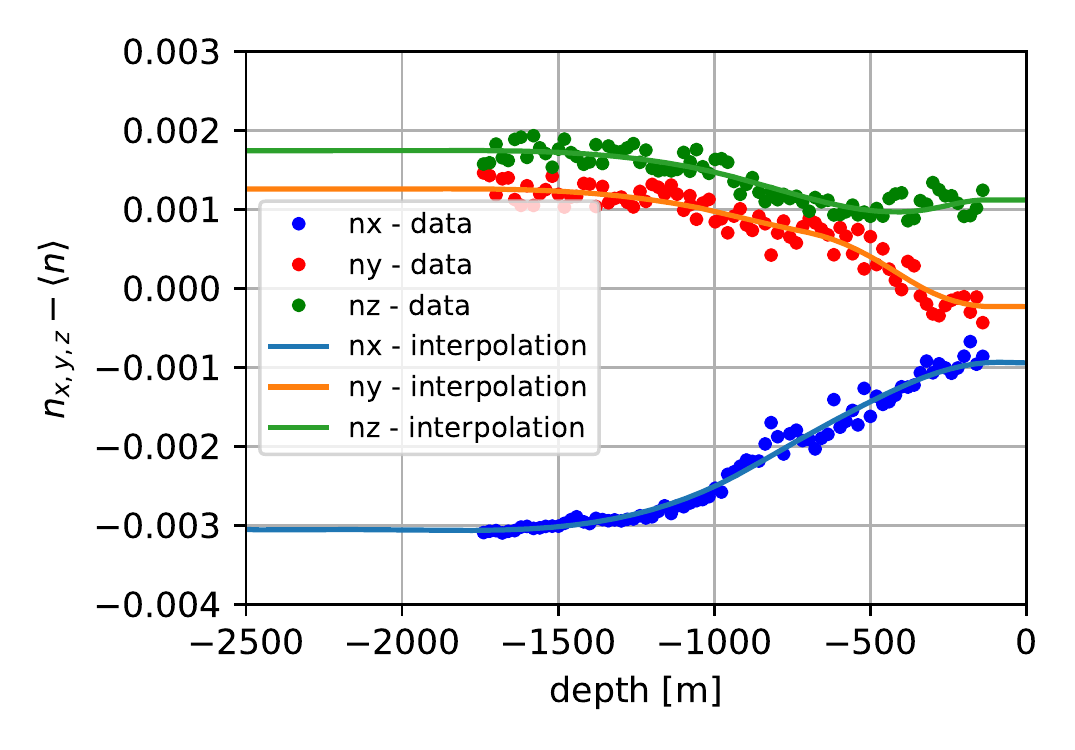}
  \caption{Refractive index as a function of depth. Measured birefringence data (scatter) \cite{voigt} compared to the average index-of-refraction value at this depth $\langle n(z) \rangle$. The shown spline interpolation of the data (solid lines) assumes constant extrapolation towards deeper and shallower depths (model A).}
  \label{fig:modelA}
\end{figure}

In addition, multiple effects such as the hexagonal crystal structure of the ice and the horizontal glacier/ice flow make the polar ice a biaxial birefringent medium. The permittivities for the different directions were measured by \cite{voigt} and the calculation on how to convert them into refractive indices can be found in \cite{Jordan}. Fig.~\ref{fig:modelA} shows this measurement as well as a spline interpolation of the data to extrapolate to deeper and more shallow depths, and to average out the presumably mostly statistical fluctuations of the measured values. As the baseline model in this article, we assume that no further change in index-of-refraction takes place towards deeper and shallower depths. We also studied three alternative choices of interpolating/extrapolating the available data that we show in the appendix in Fig.~\ref{fig:birefringence_model} which we will use later to test and verify the robustness of our birefringence predictions. 

\begin{figure}[tbp]
  \centering
  \includegraphics[ width = 0.48\textwidth]{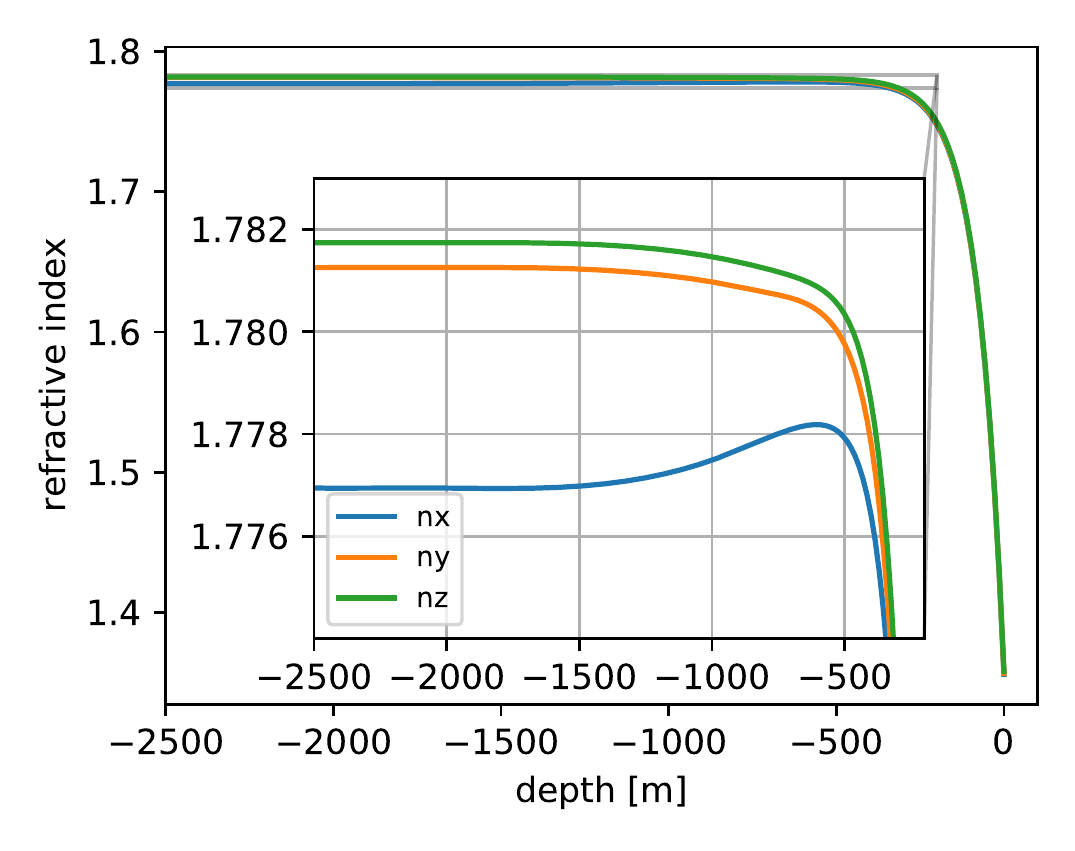}
  \caption{Refractive index as a function of depth. Combination of birefringence and density effects. The zoomed-in version highlights the birefringence effects seen in Fig.~\ref{fig:modelA} while the zoomed out version highlights the density effects from Eq.~\eqref{eq:density}.}
  \label{fig:combinationmodel}
\end{figure}

The measurement of the birefringence asymmetries is combined with the density effect to obtain a complete ice model that describes all three components of the index-of-refraction with depth $\vec{n}(z)$ which we show in Fig.~\ref{fig:combinationmodel}.

We note that the model developed here does not rely on the rather simple parameterization of density effects of Eq.~\eqref{eq:density}. Our model works for any $\vec{n}(x,y,z)$ profile. However, we will use the parameterization of Eq.~\eqref{eq:density} in the following because it generally provides good modeling of the South Pole ice \cite{southpole_2015}, it is used in current analyses \cite{ARIANNA2020Polarization,ARIANNALimit2020,ARA_limit}, and because we use the analytic ray tracer of NuRadioMC for a fast calculation of signal trajectories that only works with exponential density profiles. Typically, the parameters of the exponential index-of-refraction profile are determined from density measurements \cite{southpole_2015} which yield a good description of the bending of signal trajectories from deep in the ice to the surface as measured by the ARIANNA collaboration \cite{ARIANNA2020Polarization}. The ARA collaboration recently reported that for a propagation solely in deeper layers from \SI{200}{m} and below where the index-of-refraction is already close to the deep ice value of $n=1.78$ a modification of the parameters yield better agreement with data \cite{ARA_2022}.  The different parametrizations for ARA and ARIANNA are shown in appendix~\ref{appndix1}. In future work, we will incorporate the birefringence calculations into RadioPropa \cite{RadioPropa,RadioPropa_Git} which will allow propagation in media with arbitrary $\vec{n}(x,y,z)$ profiles. 

\subsection{Derivation of the Birefringence Model}

The following derivation was taken from \cite{light} and is repeated here to provide the relevant context. The birefringence effect can be derived from Maxwell’s equations for harmonic plane waves. 

\begin{equation} 
\begin{split}
\label{eq:plane1}
&\vec{k} \times \vec{E} = \omega \mu_0 \vec{H} \\
&\vec{k} \times \vec{H} = - \omega \epsilon \vec{E} \\
\end{split}
\end{equation}

Here, $\mu_0$ is the vacuum permeability, $\vec{k}$ is the wave vector, $\omega$ is the frequency, $\epsilon$ is the absolute permittivity, and $\vec{H}$ and $\vec{E}$ are the magnetic and electric fields of the plane wave. The general form of $\epsilon$ is a 3x3 matrix but due to our choice of the coordinate system, it reduces to a diagonal matrix. The wave equations can then be expressed via the propagation direction $\vec{s}$ (normalized), the speed of light $c$, and the relative permittivity of the medium $\epsilon_r$. 

\begin{equation} 
\label{eq:wave}
\vec{s} ( \vec{s} \cdot \vec{E}) - \vec{E} + \frac{\omega^2}{k^2 c^2} \epsilon_r \cdot \vec{E} = 0 
\end{equation}

With the effective refractive index $n = kc/\omega$ 
and using the biaxial dielectric tensor $\epsilon_r$

\begin{equation} 
\label{eq:dielectric}
\epsilon 
=
\epsilon_0 
\epsilon_r
=
 \epsilon_0 
\begin{pmatrix}
n^2_x   &   0       &   0       \\
0       &   n^2_y   &   0       \\
0       &   0       &   n^2_z   \\
\end{pmatrix} 
\end{equation}
Eq.~\eqref{eq:wave} can be written in matrix form:

\begingroup
\setlength\arraycolsep{10pt}
\begin{equation}
\label{eq:matrix}
0 = 
\begin{pmatrix}
a           &       n^2s_xs_y   &       n^2s_xs_z \\
n^2s_xs_y   &       b           &       n^2s_ys_z \\
n^2s_xs_z   &       n^2s_ys_z   &       c
\end{pmatrix}
\begin{pmatrix}
E_x \\
E_y \\
E_z \\
\end{pmatrix} 
\end{equation}
\endgroup
with
\begin{equation}
\begin{split}
a &= n_x^2 - n^2(s_y^2+s_z^2)\\
b &= n_y^2 - n^2(s_x^2+s_z^2)\\
c &= n_z^2 - n^2(s_x^2+s_y^2)\\
\end{split}
\end{equation}

Setting the determinant of the matrix in Eq.~\eqref{eq:matrix} to zero returns an equation quadratic in $n^2$ with two positive solutions which are the effective refractive indices $N_1$, $N_2$. 

\begin{equation} 
\begin{split}
\label{eq:n}
&\left(n_x^2 - n^2\right)\left(n_y^2 - n^2\right)\left(n_z^2 - n^2\right) \\ 
& + n^2 \big[ s_x^2 \left(n_y^2 - n^2\right)\left(n_z^2 - n^2\right) + s_y^2 \left(n_x^2 - n^2\right)\left(n_z^2 - n^2\right) \\
&+ s_z^2 \left(n_x^2 - n^2\right)\left(n_y^2 - n^2\right)\big] = 0
\end{split}
\end{equation}

The roots can be found analytically using the equations in~\ref{appndix2} which were implemented into NuRadioMC. The two corresponding polarization eigenvectors can be calculated via 

\begin{equation} 
\label{eq:pol}
\vec{e_i} = 
\begin{pmatrix}
\dfrac{s_x}{N_i^2 - n_x^2} \\
\dfrac{s_y}{N_i^2 - n_y^2} \\
\dfrac{s_z}{N_i^2 - n_z^2} \\
\end{pmatrix}
\end{equation}
with $i = 1,2$.
For special geometries where $N_i = n_{x,y,z}$, Eq.~\eqref{eq:n} simplifies, and these special cases are treated separately \cite{Nils}. The two eigenvectors are orthogonal to each other. It is convenient to express the resulting vector in spherical coordinates $\theta$, $\phi$, and $r$, where the $r$-component can be neglected as it is close to zero. Normally, electromagnetic waves are polarized orthogonal to their direction of propagation resulting in a zero $r$-component but in a dielectric medium, the Poynting vector can deviate from the propagation direction to which the wave is orthogonal. In the case of polar ice, due to the small birefringence asymmetries, the $r$ component is not exactly zero but at the level of a few permil compared to the amplitude of the $\theta$ and $\phi$ components which we ignore in the following. 

The derivation shows that a radio wave splits up into two orthogonal components with two different effective indices-of-refraction that depend on the propagation direction $\vec{s}$ and the dielectric tensor $\epsilon$ where $\epsilon$ reduces to the refractive index vector $\vec{n}$ due to our choice of coordinate system that orthogonalizes the tensor. 

\subsection{Pulse Propagation Model}
Both, the propagation direction and the refractive index change during the propagation of the radio signal. We account for that by performing the propagation in small incremental steps over which these quantities can be assumed to be constant. As a first step, the signal trajectory through the ice is calculated ignoring birefringence effects. In this work, we use the analytic ray tracer of NuRadioMC \cite{NuRadioMC} but also other propagation codes that support more complex index-of-refraction profiles and deviate from the exponential modeling of Eq.~\eqref{eq:density} could be used. Work to integrate this model into the numerical ray tracer RadioPropa \cite{RadioPropa,RadioPropa_Git} is ongoing. 

We use the same propagation direction for both propagation states at each incremental step. This neglects the small spatial separation of the two states in the firn due to the small difference in effective index-of-refraction and therefore slightly different propagation paths. We think that this approximation is justified and discuss it in Sec.~\ref{sec:approx}.

By subtracting the propagation time calculated from $N_1$ with the propagation time calculated from $N_2$, Eq.~\eqref{eq:time} can be used to calculate the time delay $\Delta T$ due to birefringence. 

\begin{equation} 
\label{eq:time}
\Delta T = \frac{l}{c} (N_1-N_2)
\end{equation}

\begin{figure*}[tbp]
  \centering
  \includegraphics[width = 0.7\textwidth]{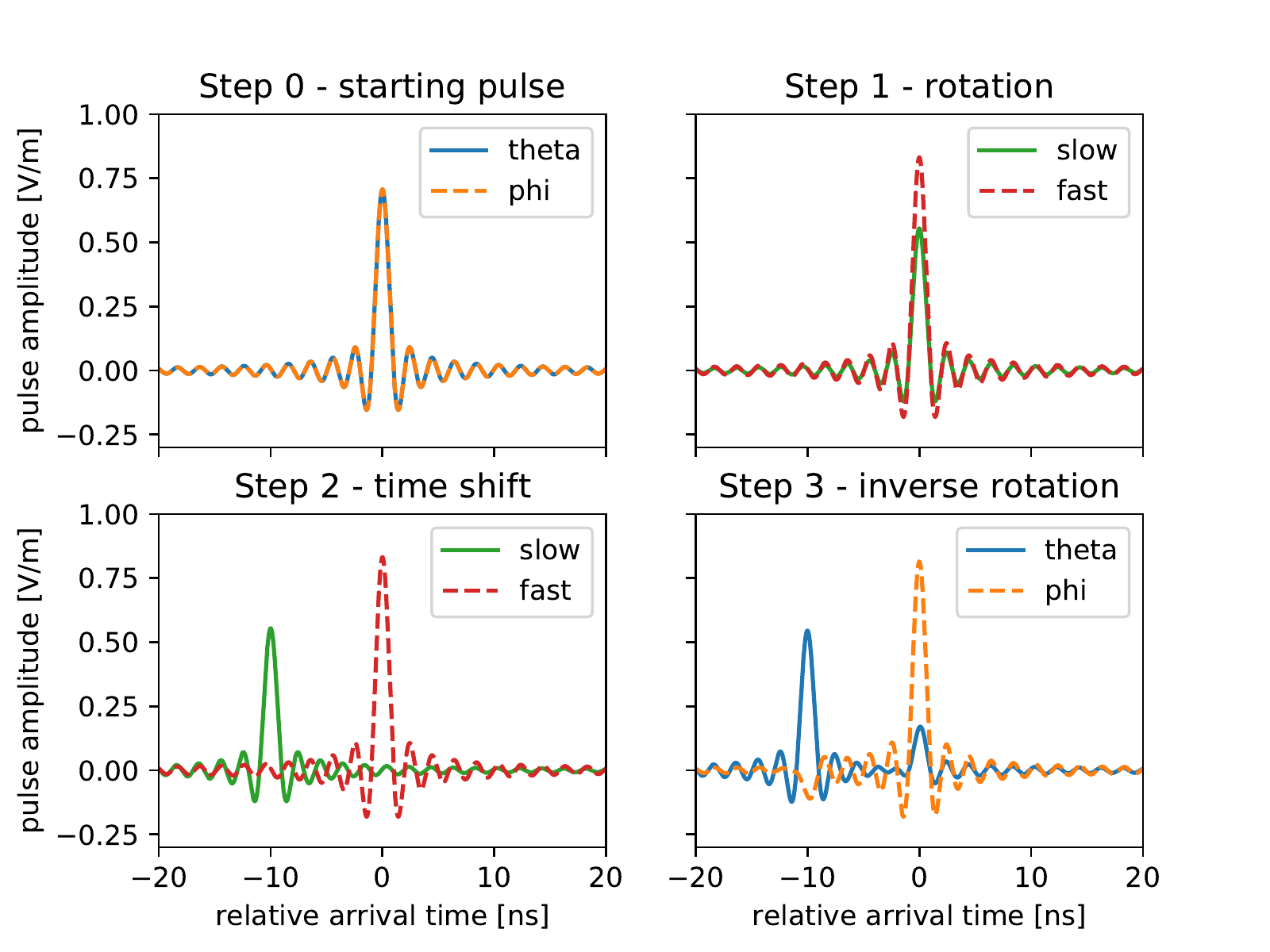}
  \caption{Pulse propagation model for an artificial pulse shape with the created time traces at the source (step 0), the rotation into the time domain (step 1), the time shift applied to separate the pulses (step 2), and the inverse rotation back into the polarization states (step 3).}
  \label{fig:prop_model}
\end{figure*}

As the polarization eigenvectors of the two propagation states change while the pulse propagates (due to a change in $\vec{n}$ as well as a change in the direction of propagation), a mixing of fast and slow parts of the pulse occurs. This means that a simple calculation of time differences between the two propagation states using Eq.~\eqref{eq:time} as was done in previous work \cite{Jordan,Amy} is not sufficient to properly describe the effect of birefringence for the radio detection of neutrinos. 

To account for the change in the eigenvectors, a pulse propagation model for arbitrary waveforms was created which we illustrate in Fig.~\ref{fig:prop_model}. The polarization of an electromagnetic wave is best described in spherical coordinates where the radial component is always zero due to the transversality of the wave. Then, any waveform can be described by specifying the pulse shape in the \emph{theta/phi} basis (step 0). This pulse is then rotated into the new \emph{slow/fast} basis by a rotation matrix $R$ defined by the two eigenvectors $\vec{e}_{1,2}$ calculated from Eq.~\eqref{eq:pol} where one part of the pulse travels faster than the other (step 1). The incremental time shift calculated from Eq.~\eqref{eq:time} is then applied to the pulses (step 2) and the pulses are rotated back into the natural theta and phi states (step 3). This process is illustrated in Fig.~\ref{fig:prop_model} and described in the following equations.

The rotation matrix is given by the unit vectors of the two polarization eigenvectors (and the one orthogonal to them to form a new orthogonal basis)
\begin{equation}
\label{eq:R}
R = 
\begin{pmatrix}
e_\perp^r  &     e_\perp^\Theta &   e_\perp^\Phi   \\
e_{1}^r         &     e_1^\Theta        &   e_1^\Phi            \\
e_{2}^r         &     e_2^\Theta        &   e_2^\Phi            \\
\end{pmatrix}
\approx
\begin{pmatrix}
1   &     0             &   0   \\
0   &     e_1^\Theta    &   e_1^\Phi            \\
0   &     e_2^\Theta    &   e_2^\Phi            \\
\end{pmatrix}
\end{equation}

As shown in equation \ref{eq:R}, the three-dimensional problem can be reduced to a two-dimensional problem as the radial component of the pulse polarization is almost zero. Then the waveform described as the amplitude as a function of time $A(t)$ transforms as follows
 
\begin{equation}
\label{eq:rot}
\begin{pmatrix}
A_{N_1}(t)        \\
A_{N_2}(t)        \\
\end{pmatrix}
=
R
\begin{pmatrix}
A_{\Theta}(t)        \\
A_{\Phi}(t)        \\
\end{pmatrix}
\end{equation}

Then the time shift is applied via

\begin{equation}
\label{eq:trans}
A_{N_1}(t) \rightarrow A_{N_1}(t - \Delta t)
\end{equation}

We implement the time shift using the Fourier shift theorem, i.e., that a translation in the time domain corresponds to a multiplication by a phase factor in Fourier domain. This allows to shift waveforms precisely even with $\Delta t's$ that are smaller than the binning of the time domain. To increase performance, the complete propagation is performed in the Fourier domain. Finally, the waveforms are rotated back into the $\theta$/$\phi$ basis. 

\begin{equation}
\label{eq:rot_back}
\begin{pmatrix}
A_{\Theta}(t)        \\
A_{\Phi}(t)        \\
\end{pmatrix}
=
R^{-1}
\begin{pmatrix}
A_{N_1}(t)        \\
A_{N_2}(t)        \\
\end{pmatrix}
\end{equation}

\begin{figure*}[tbp]
  \centering
  \includegraphics[width = 0.7\textwidth]{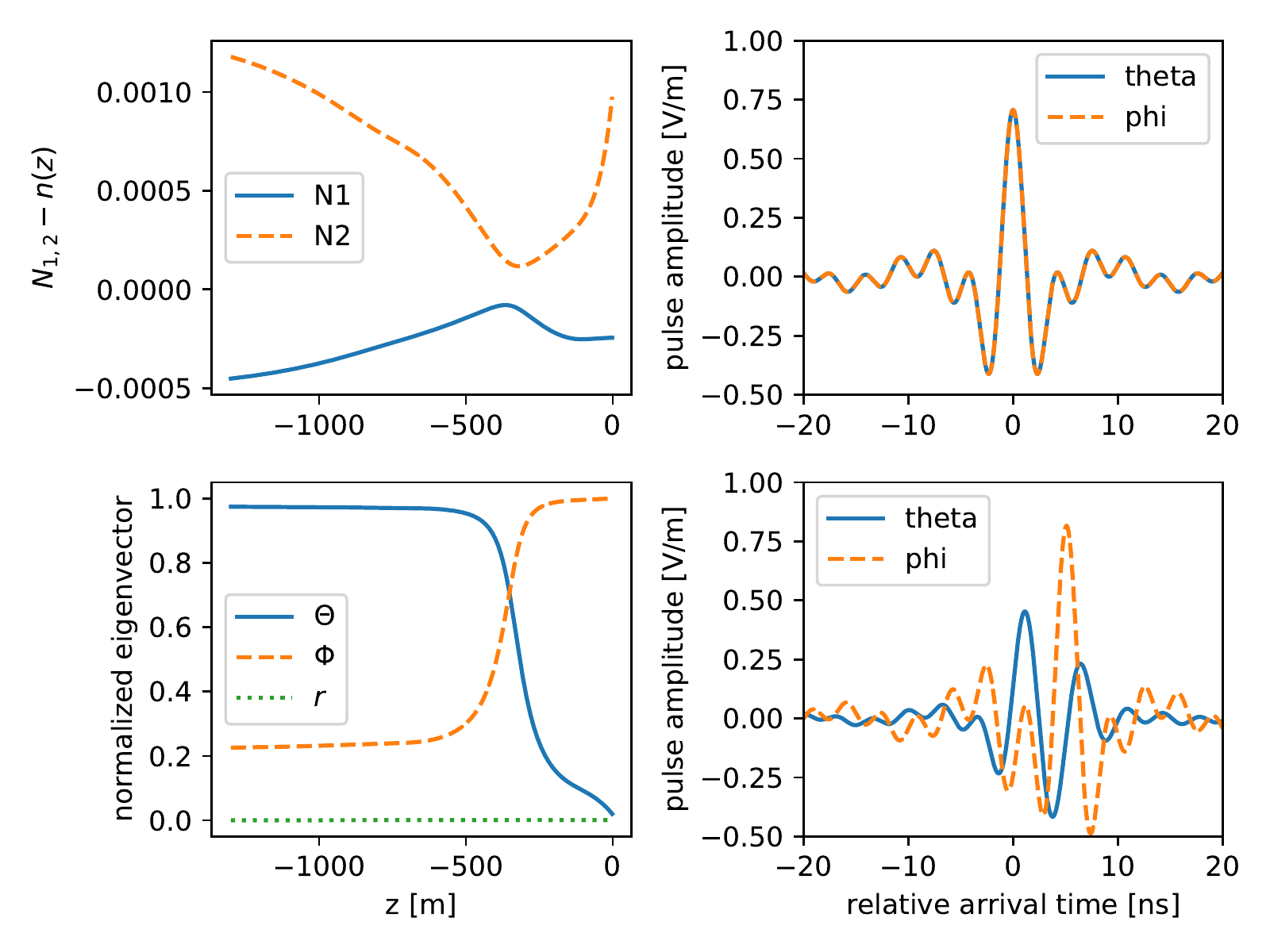}
  \caption{Display of the pulse propagation properties with vertex position at [\SI{0}{m}, \SI{0}{m}, \SI{-1300}{m}] and antenna position at [\SI{1500}{m}, \SI{200}{m}, \SI{-1}{m}]. Top left: Effective refractive indices $N_{1,2}$ against depth. The density effect is subtracted to be able to see the relative behaviour. Bottom left: Normalized eigenvector of $N_1$ against depth. $N_2$ behaves the same way with $\Theta$ and $\Phi$ switched. Top right: Waveform before propagation. Bottom right: Waveform after propagating.}
  \label{fig:first_ex}
\end{figure*}

Steps 1-3 are then repeated for every incremental step of the signal propagation.

We show an example of a pulse propagating from \SI{1300}{m} depth to a receiver close to the surface at a horizontal distance of \SI{1500}{m} at an angle that is close to parallel to the ice flow in Fig.~\ref{fig:first_ex}. We only show the direct trajectory to the receiver. We choose a generic bandpass limited delta pulse with equal amplitude in the $\theta$ and $\phi$ polarization states as starting pulse. The example shows how the two effective indices of refraction as well as how the polarization eigenvector change during propagation. We picked a geometry where the change in eigenvectors together with the accumulation of time delays lead to interference which is visible in the pulse shapes after propagation. 

\subsection{Limitations of the model and future cross-checks}
\label{sec:approx}
Because we derived the calculation directly from first principles, the model does not have any free parameters that can be tuned. The only input is the index-of-refraction profile and the initial waveforms before propagation. However, we assume that the propagation can be described with ray optics and that both eigenstates propagate along the same path. The validity of this assumption is difficult to verify without doing a much more elaborate calculation. Even if we would keep the ray tracing assumption but would take into account path differences due to the small differences in the index-of-refraction, the complexity of the simulation would increase dramatically because, in every propagation step, the path would split in two resulting in an exponential increase in paths to consider of $2^{N_\mathrm{steps}}$ which would lead to $10^{301}$ path segments for a typical number of 1000 propagation steps. We approximate the resulting uncertainty by considering the most extreme case of calculating the difference between the propagation paths of two pulses that following density profiles that differ by the maximal observed difference between $n_x$, $n_y$, and $n_z$ of $n_{ice} = 0.05$. The difference is typically only a few mm and always stayed below \SI{0.5}{m} which was only reached during a small part of the propagation. Because the path differences are small and in particular smaller than the considered wavelengths of $\sim$\SI{0.5}{m} - \SI{1.5}{m} we think that the approximation is justified. 

There is a way to test our calculations via Finite Difference Time Domain (FDTD) simulations that essentially evolves Maxwell's equations over time within some finite computational volume, e.g., using the open-source code MEEP \cite{meep2010}. FDTD simulations have already been used to study second-order propagation effects for radio waves in polar ice \cite{Deaconu2018}. However, FDTD simulations are extremely CPU and memory intensive, especially for the large volumes $\mathcal{O}$(\SI{1}{km^3}) and high frequencies relevant for us. About 10 grid points are needed per wavelength. Even restricting the highest frequency to \SI{200}{MHz} and using $n=1.78$ resulting in $\lambda$ = \SI{84}{cm} will result in one grid point every $\SI{7}{cm}$. Then, the memory consumption can easily reach several TB and the computation time would exceed 100k core hours. However, using a supercomputer where several large computing nodes are combined using MPI, such a simulation seems in principle possible. We plan to consider this in future work. However, due to the large computing costs, such a cross-check could only be done for a few selected geometries. For all practical purposes, a fast model as presented here is required. 

\section{SPICE measurement setup}

\begin{figure}[tbp]
  \centering
  \includegraphics[width = .48\textwidth]{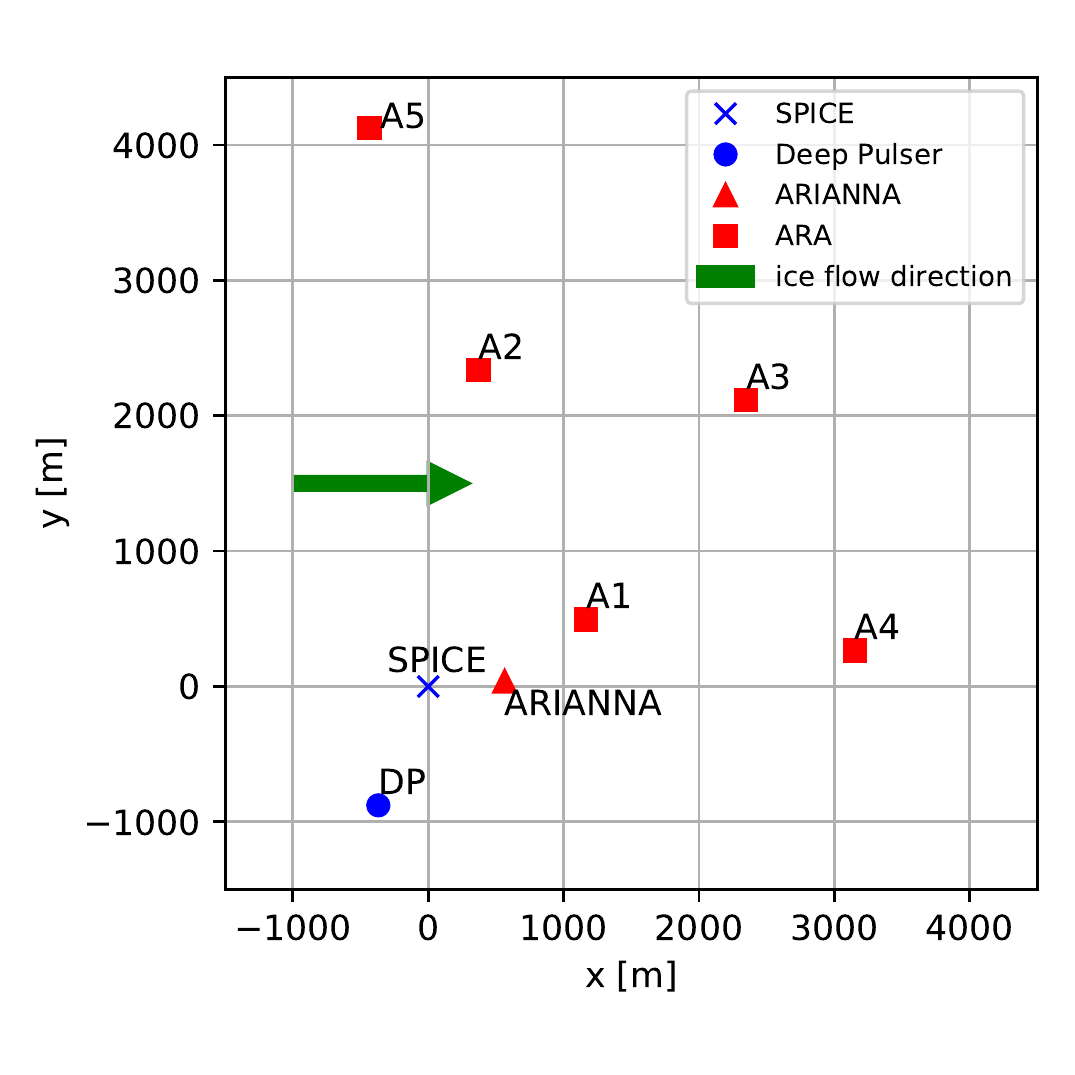}
  \caption{The geometry of the measurement setup at the South Pole \cite{Amy}, with the SPICE hole for the transmitter in blue and the ARA and ARIANNA antenna stations in red. Also shown is the position of the \emph{deep pulser} \cite{Jordan}, another radio transmitter attached to the end of one of the IceCube strings.}
  \label{fig:geom_bird}
\end{figure}

Next to its original purpose of measuring the ice properties of the South Pole, the SPICE borehole can now be used to calibrate radio detector stations and to study the propagation of radio waves through polar ice. To do this, a radio transmitter, repeatedly emitting short radio pulses, was lowered down the borehole. Radio antennas - placed into the ice at shallow depths - then measured the pulses after in-ice propagation. In the following, we use public data from measurement campaigns done by the ARA and ARIANNA collaborations \cite{ARIANNA2020Polarization,Allison2019,Jordan} to which we compare our predictions of birefringence effects. The positions of the different ARA and ARIANNA radio detector stations are indicated in Fig.~\ref{fig:geom_bird} with respect to the SPICE hole at (0, 0)  and the ice flow in the x-direction. 

The ARA detector stations consist of antennas installed to a depth of down to \SI{200}{m}. Due to the limited diameter of the borehole, cylindrical bicone antennas (that are sensitive to the vertical polarization component, named vpol in the following) and quad-slot antennas (that are sensitive to the horizontal component, named hpol in the following) are used. Each station consists of four strings. Eight pairs of vpol and hpol antennas form a cube with a separation of approx. \SI{20}{m} \cite{ALLISON2012457} with slight variations from station to station. 

The ARIANNA detector station consists of four LPDA antennas that point downwards and are buried approx.~\SI{1}{m} below the snow surface. They are arranged horizontally in two parallel pairs that are orthogonal to each other. This setup allows for the reconstruction of the three-dimensional electric field via a simultaneous unfolding of the antenna response of the four LPDAs. The measurement setup and the corresponding analysis are described in detail in Ref.~\cite{ARIANNA2020Polarization}. 

\begin{figure}[tbp]
  \centering
  \includegraphics[ width = .48\textwidth]{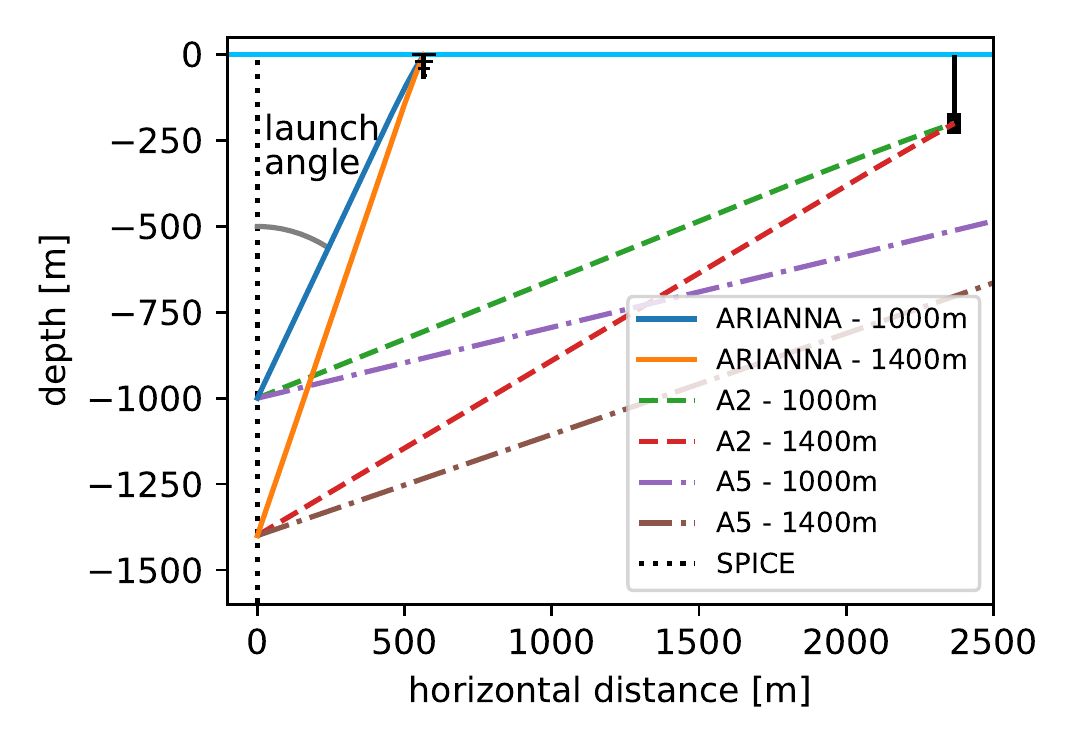}
  \caption{Side view of two radio signals propagating from the transmitter in the SPICE hole at \SI{-1000}{m} and \SI{-1400}{m} depth to the antennas of the ARA and ARIANNA stations. Detector station ARA5 is outside of the plot range but the signal trajectories are shown. }
  \label{fig:geom_side}
\end{figure}

Examples of direct signal trajectories from two representative emitter depths are shown in Fig.~\ref{fig:geom_side} with the ARIANNA antenna sitting at $\sim$\SI{1}{m} below the snow surface and the ARA receivers going down to $\sim$\SI{200}{m}.

The emitter that was lowered into the SPICE borehole consisted of a pulse generator that was connected to a fat dipole antenna. It was measured carefully in an anechoic chamber to have a precise model of the emitted pulses. The measurement setup and results are described in detail in \cite{ARIANNA2020Polarization} and are available at \cite{G_Git}. The pulses were measured for different launch angles, i.e., the angle between the line-of-sight from the emitter to the receiver and the symmetry axis of the dipole antenna. In the measurement setup at the South Pole, the launch angle changes with the depth of the transmitter in the SPICE borehole (cf. Fig.~\ref{fig:geom_side}). Naively one would expect that the emitted signal is only $\theta$ polarized but the measurement revealed the presence of a cross-polarization ($\phi$ polarization) amplitude of up to 20\% depending on the launch angle. 

\begin{figure}[tbp]
  \centering
  \includegraphics[width = 0.48\textwidth]{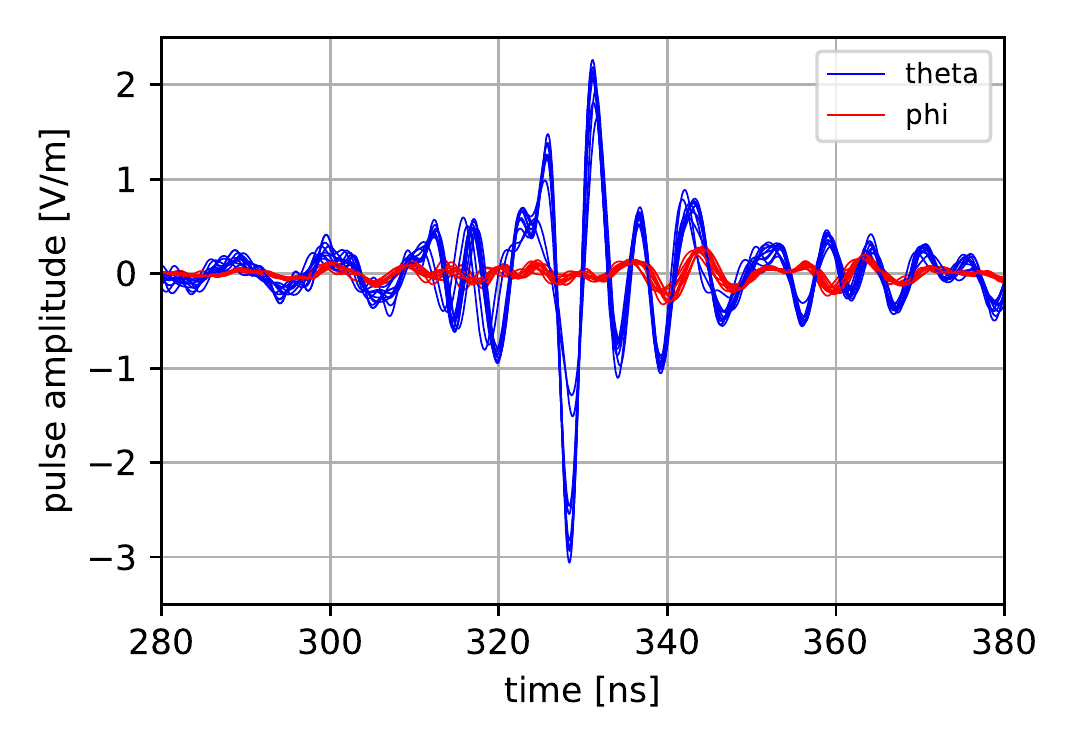}
  \caption{Pulses emitted at a launch angle of $30^\circ$ from the anechoic chamber. Theta traces are indicated in blue and phi traces are indicated in red.}
  \label{fig:pulse_all}
\end{figure}

The electric field was measured for launch angles of $15^\circ$, $30^\circ$, $45^\circ$, $60^\circ$, $75^\circ$ and $90^\circ$. Each measurement was repeated 10 times. In Fig.~\ref{fig:pulse_all}, we show the set of 10 measurements for a launch angle of \SI{30}{\degree}. Apart from the existence of a non-zero cross-polarization amplitude, one can also see a relatively large variance of the emitted pulses for the same launch angle. Not only does the amplitude vary but the theta and phi components have their maximum at different times which further complicates the interpretation of time delay measurements. Our pulse propagation code allows, for the first time, to take these subtle but important effects into account. 

It is often useful to reduce the information of the electric-field traces to a single quantity. A common choice that was used in previous work \cite{ARIANNA2020Polarization} is to calculate fluence and polarization. The fluence is calculated by integrating the squared electric field. We use the same integration window as used in Ref.~\cite{ARIANNA2020Polarization} of \SI{70}{ns} around the maximum of the dominant $\theta$ component, and subtract the contribution of noise: 

\begin{equation} 
\label{eq:F}
f_{\theta,\phi} = \sqrt{\sum_{t = t_m -35ns}^{t_m +35ns} |E_{\theta,\phi}(t)|^2} - f_{\theta,\phi, noise} \,.
\end{equation}

This information can be further reduced to a single polarization angle per electric-field pulse which is given by:

\begin{equation} 
\label{eq:P}
P = \arctan \left( \frac{f_\phi}{f_\theta} \right) \,.
\end{equation}

\begin{figure}[tbp]
  \centering
  \includegraphics[width = 0.5\textwidth]{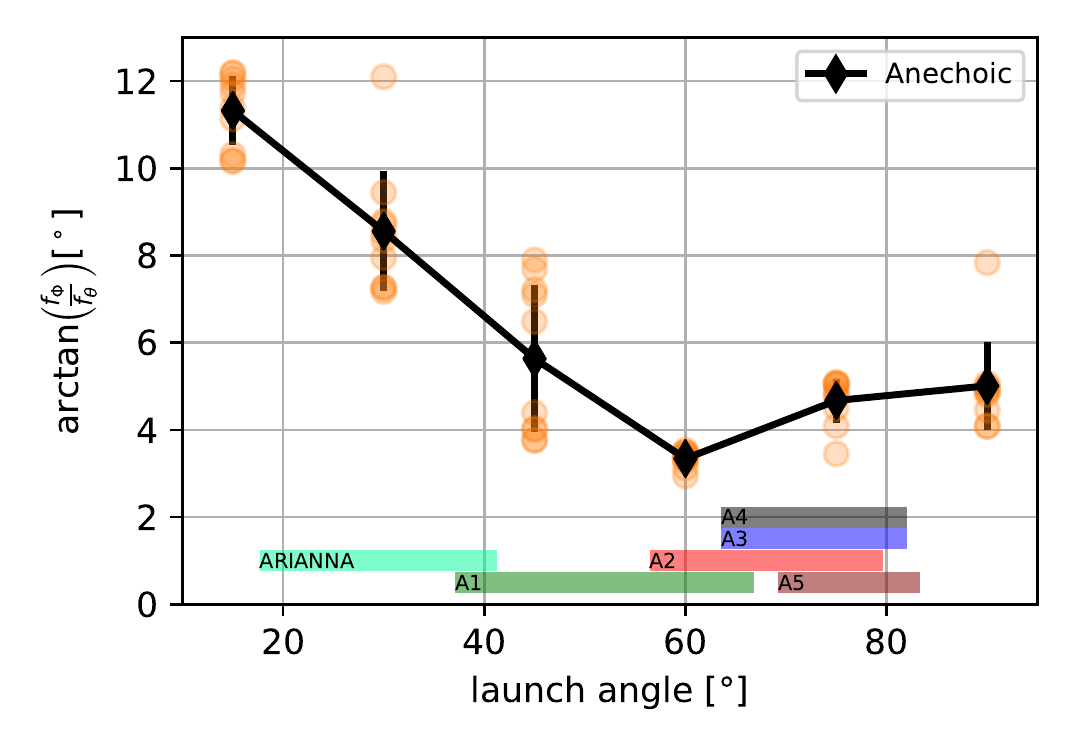}
  \caption{Polarization of starting pulses against the launch angle. The launch angle at which each station can receive signals from the SPICE drop is indicated below. Figure adapted from Ref.~\cite{ARIANNA2020Polarization}.}
  \label{fig:pol_vs_launch}
\end{figure}

The polarization of the emitted pulses as a function of the launch angle is shown in Fig.~\ref{fig:pol_vs_launch} as well as which launch angles are covered for the different geometries and transmitter depths from \SI{600}{m} to \SI{1700}{m}.

\section{Comparison to ARA data}

As seen in Fig.~\ref{fig:geom_bird}, the ARA collaboration installed multiple detector stations at the South Pole with various angles to the ice flow relative to the direction to the SPICE hole. This gives the ARA experiment many useful handles for measuring the effect of birefringence. Previous studies measured time delays between signal pulses arriving at the vpol and hpol antennas \cite{Jordan}. This measurement of polarization-dependent time delays is a direct test of birefringence. In previous work using a simplified model of birefringence, the measured time delays could be explained for some of the geometries but showed deviations for other geometries \cite{Jordan}.

In the following, we make a detailed prediction of the ARA measurements by propagating the emitted waveforms (obtained from the anechoic chamber measurements described above) to the receivers using the birefringence model described in Sec.~\ref{sec:birefringence_model}.

\subsection{Time delay}

\begin{figure*}[tbp]
  \centering
  \includegraphics[width = .75\textwidth]{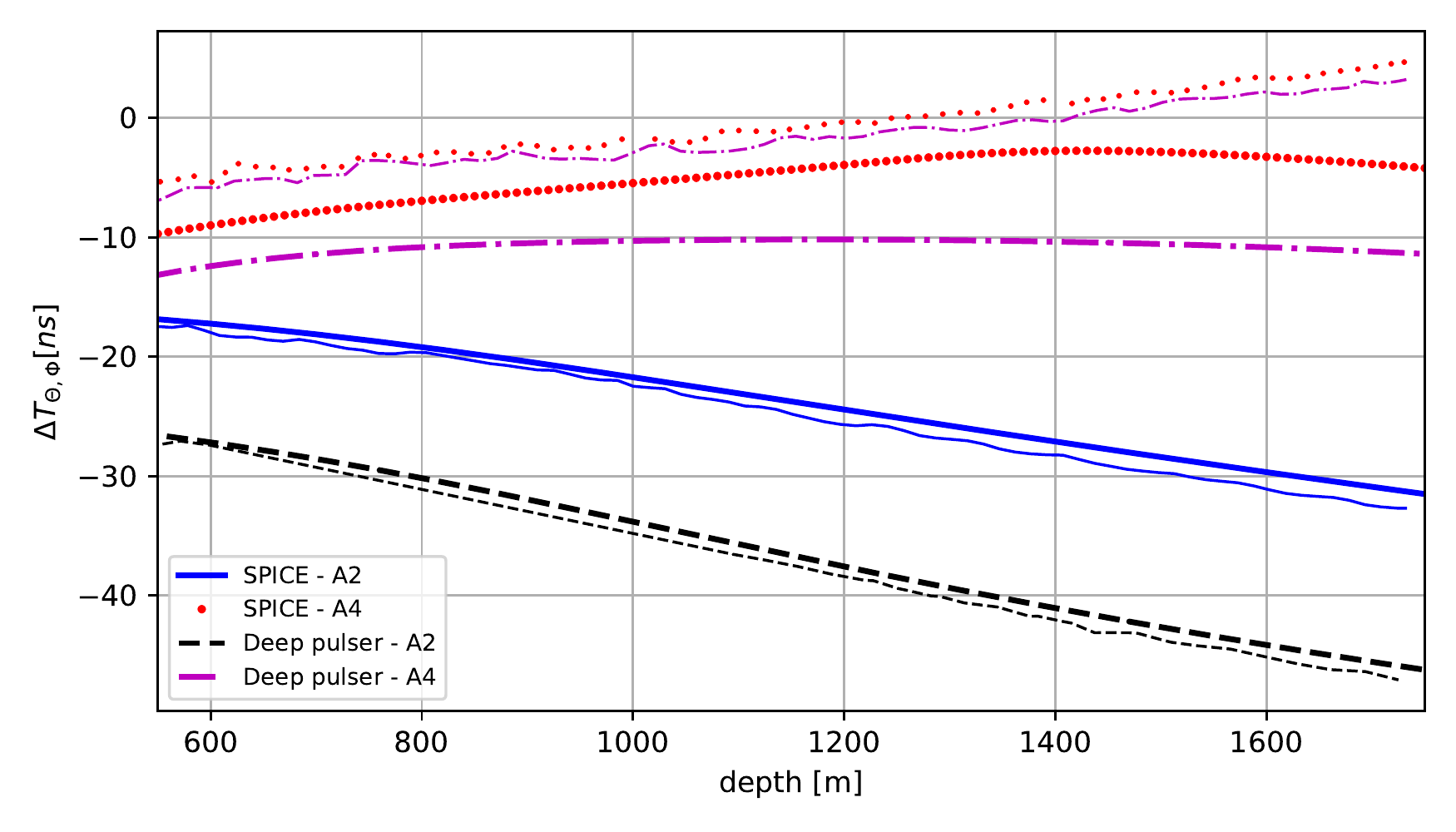}
  \caption{The simulated time delay from \cite{Jordan} (thin lines) compared to the predictions from the introduced model (thick lines).}
  \label{fig:Jordan_comp}
\end{figure*}

Simulating the time delay is difficult, as many small effects can change the outcome of a measurement. To make a comparable prediction to the previous simulation of \cite{Jordan}, the time delay was first calculated independently of the waveforms by adding the incremental time delays between the $\theta$ and $\phi$ polarization components resulting from equation~\ref{eq:time}. The results are shown in Fig.~\ref{fig:Jordan_comp}.

For the geometries perpendicular to the ice flow, our predictions are very similar to the previous calculation. For parallel geometries, one can see a similar trend but a larger discrepancy to the previous model. The discrepancy can be explained when considering how the two models calculate the effective refractive indices. The previous model from \cite{Jordan} approximated a perfect propagation in the x-z-plane for the parallel case and a perfect propagation in the y-z-plane for the perpendicular case. Then, one can approximate $N_1$ with $n_x$ ($n_y$) and $N_2$ with a weighted average of $n_y$ and $n_z$ ($n_x$ and $n_z$) for a perfectly parallel (perpendicular) propagation with respect to the ice flow. However, the geometries for the A2 and A4 station are not perfectly perpendicular or parallel to the ice flow and so $N_1$ and $N_2$ both depend $n_x$, $n_y$ and $n_z$. Our model takes this into account. Due to $n_z$ and $n_y$ having similar values the approximation works better for the perpendicular case (propagation to ARA station A2) as $N_2$ lies in between $n_z$ and $n_y$. For the parallel case (propagation to ARA station A4) $N_2$ lies between $n_z$ and $n_x$ where the room for error is larger and this difference becomes apparent in Fig.~\ref{fig:Jordan_comp}. Other notable differences between the models include a larger numerical step size ($\Delta\sim \SI{20}{m}$) and using the exact data points of Fig.~\ref{fig:modelA} for the model described in \cite{Jordan} compared to $\Delta\sim\SI{2}{m}$ and the interpolation of the data points shown in section~\ref{section: Ice Model} for our model.

When comparing the simulated predictions to measured data one has to be careful to distinguish between the natural polarization states (here theta and phi) and what the antennas measure (vpol and hpol). The vpol antenna is only sensitive to (the vertical projection of) the theta component of the electric field. However, the hpol antenna is sensitive to theta as well as the phi component because the theta component also has a horizontal component depending on the signal direction. Only for horizontal signal directions, the theta component is purely vertical. In our analysis, we calculate the response of both antennas using a detailed model of the antenna response that is available through NuRadioReco \cite{NuRadioReco}. 

The ARA collaboration presented one measurement each for A2 and A4 \cite{Jordan} of the time delay between the pulse amplitude in the vpol and hpol antennas for an emitter depth of \SI{1000}{m} with $\Delta T_{v, h}(\SI{1000}{m}, A2) = -14.1 \pm \SI{2.8}{\nano \second}$ and $\Delta T_{v, h}(\SI{1000}{m}, A4) = 4.6 \pm \SI{9}{\nano \second}$. 

To make a thorough prediction for these data points, we start with the waveforms from the anechoic chamber measurement at $75^\circ$ launch angle that matches best the geometry (cf. Fig.~\ref{fig:pol_vs_launch}). One of the ten measured waveforms showed a different behavior than the others which we attribute to measurement error and therefore disregard this waveform.
Then, we  propagate the remaining 9 waveforms to the receiver using our birefringence code, correct for attenuation and fold  with the antenna response to obtain the expected voltage traces in the vpol and hpol antennas. 
We calculate the time difference between the maxima of the Hilbert envelopes of these components from which we calculate the time delay. We use the standard deviation of the calculated time delays of the 9 pulses to estimate the uncertainty of the prediction. This way, also the initial time difference between the pulses is accounted for,  a crucial aspect that was neglected in previous analyses. 

The results are summarized in table \ref{tab: timeD}.  For both geometries our predictions agree with the measurements within uncertainties, with a difference of $\sim \SI{1}{\nano \second}$ for the A4 geometry and $\sim \SI{3}{\nano \second}$ for the A2 geometry. The application of the antenna response ($\Delta T_{\Theta, \Phi}$ vs. $\Delta T_{v, h}$) constitutes a minor correction because the signal arrival direction is close to horizontal. However, it is crucial to propagate the emitted waveforms compared to the simplified calculation done previously which leads to differences of up to \SI{10}{ns}. Furthermore, our more precise calculation of the effective indices-of-refraction makes a significant difference for the A4 geometry compared to the approximation done in \cite{Jordan}. 

The agreement of our prediction with the measured data points indicates that our model describes the South Pole ice, and that time delay measurements are useful to probe birefringence effects.  However, these are only two data points for which a comparison could be made and further comparisons to real data are needed to solidify the prediction power.

\begin{table*}
\begin{center}
\begin{tabular}{c|c c | c c c}
  & \multicolumn{2}{c|}{simple calculation} & \multicolumn{2}{c}{pulse propagation} & measured data\\
 \hline
  & \bm{$\Delta T_{\Theta, \Phi}$} \textbf{[\si[detect-weight]{\nano \second}]} & $\Delta T_{\Theta, \Phi}[\si{\nano \second}]$ & \bm{$\Delta T_{\Theta, \Phi}$} \textbf{[\si[detect-weight]{\nano \second}]} & \bm{$\Delta T_{v, h}$} \textbf{[\si[detect-weight]{\nano \second}]} & $\Delta T_{v, h}[\si{\nano \second}]$ \\ 
 & (this work) & (Ref.~\cite{Jordan}) & (this work) & (this work) & (ARA \cite{Jordan})  \\ 
\hline \hline
  SPICE - A2 & \bm{$-21.7$} & $-22.5$ & \bm{$-11.6 \pm 0.7$} &  \bm{$-10.9 \pm 0.4$} & $-14.1 \pm 2.8$ \\ 

  SPICE - A4 & \bm{$-5.5$} &  $-1.6$ & \bm{$4.1 \pm 1.1$} &  \bm{$3.9 \pm 0.2$} & $4.6 \pm 9$ \\ 
\end{tabular}
 \caption{Summary of time delay measurements and predictions for the pulse propagation \SI{1000}{m} depth to the ARA A2 and A4 stations. The first two columns show the accumulated time delay in the theta and phi polarization component using Eq.~\eqref{eq:time} from our model and the simplified calculation from \cite{Jordan}. The third/fourth columns show the time delay extracted from the propagated waveforms of the electric field (theta, phi) and include the antenna response (v, h). The last column shows the measured time delay between the vpol and hpol antennas. See text for details. The values indicated in bold fond represent our predictions and the normal fond values represent the predictions and measurements published in \cite{Jordan}.}
 \label{tab: timeD}
 \end{center}
\end{table*}

\subsection{Amplitude}

\begin{figure}[tbp]
  \centering
  \includegraphics[width = 0.48\textwidth]{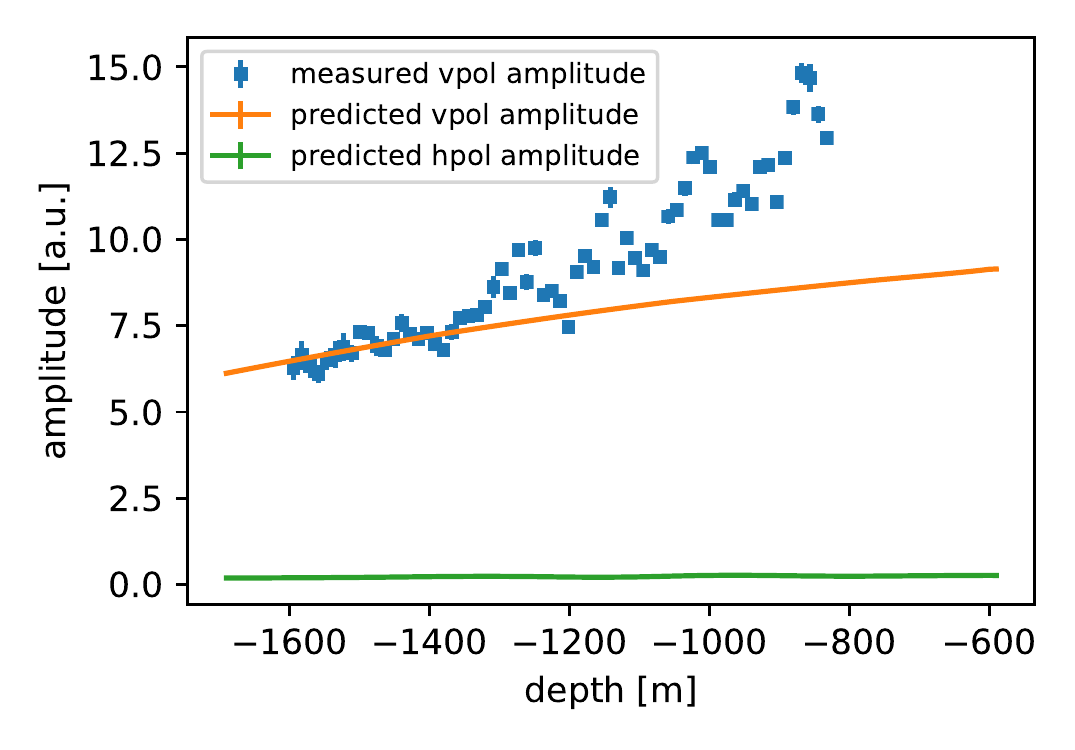}
\caption{The amplitude of the vpol and hpol component of the pulses reaching the ARA A5 station from different depths. The measured vpol data is indicated in blue and the normalized vpol predictions from the birefringence model are indicated in orange. ARA datapoints are published in \cite{Allison_2020}.}
\label{fig:A5}
\end{figure}

Another potential way to experimentally probe birefringence effects is via changes in the pulse amplitude or polarization with emitter depth. As seen in Fig.~\ref{fig:first_ex}, the polarization and/or pulse amplitude can change due to interference because the birefringence polarization eigenvectors are different from the $\theta$ and $\phi$ states and change during propagation. We note that this effect is only pronounced for geometries where the resulting time delays are small because we are only concerned with measuring short, few-nanoseconds-long pulses.

The ARA collaboration presented a measurement of the vpol signal strength (averaged over all 8 vpol antennas of the station) where they saw an oscillatory behavior in the pulse amplitude with emitter depth \cite{Allison_2020} which we show in Fig.~\ref{fig:A5}. It was hypothesized that birefringence was the cause of this amplitude variation \cite{Amy}. Unfortunately, the data quality did not allow a reconstruction of the electric field (and thereby signal polarization) to probe experimentally that the change in amplitude originated from a change in signal polarization which would be the signature of birefringence. The birefringence model presented in \cite{Amy} was able to qualitatively generate such amplitude oscillations. It assumed continuous waves of a fixed frequency. With this assumption, as the time delay increases with propagation length, i.e., emitter depth, it leads to an interference pattern as a function of depth. 

To investigate this behavior using more realistic conditions, we use waveforms similar to Fig.~\ref{fig:pulse_all} but again for a launch angle of $75^\circ$ to match the geometry. We propagate these waveforms from the SPICE hole at different depths to the A5 station. The resulting waveforms were corrected for attenuation and antenna response effects and the amplitudes were determined by calculating the maximum of the Hilbert envelope of the vpol and hpol signals. The amplitudes were then rescaled to the measured amplitudes in order to account for the amplification in the ARA signal chain which is unknown to us. We plot the resulting amplitudes against the depth of the transmitter and compare it to the measured data in Fig.~\ref{fig:A5}. We fail to generate any oscillations due to birefringence.  We attribute this to the fact that we use realistic pulse shapes and not the unrealistic assumption of continuous waves. 
We speculate that some part of the observed effect might originate from a combination of averaging over the 8 vpol antennas that are at different depths and an increase of signal amplitude when the shadow zone boundary is approached by the transmitter due to a focusing effect \cite{NuRadioMC}. 

\section{Comparison to ARIANNA data}

The ARIANNA collaboration aims to measure neutrinos with shallow radio stations \cite{ARIANNALimit2020}. Most detector stations are installed on the Ross Ice shelf but two ARIANNA stations were installed at the South Pole where one of them was close enough to the SPICE hole to be able to observe the emitted signals from the SPICE pulser drop. The ARIANNA collaboration was able to reconstruct the signal direction with sub-degree precision as well as the electric field pulse that arrived at the detector station \cite{ARIANNA2020Polarization}. This allows a detailed measurement of the signal pulse properties as a function of emitter depth. 

\begin{figure*}[tbp]
  \centering
  \includegraphics[width = .9\textwidth]{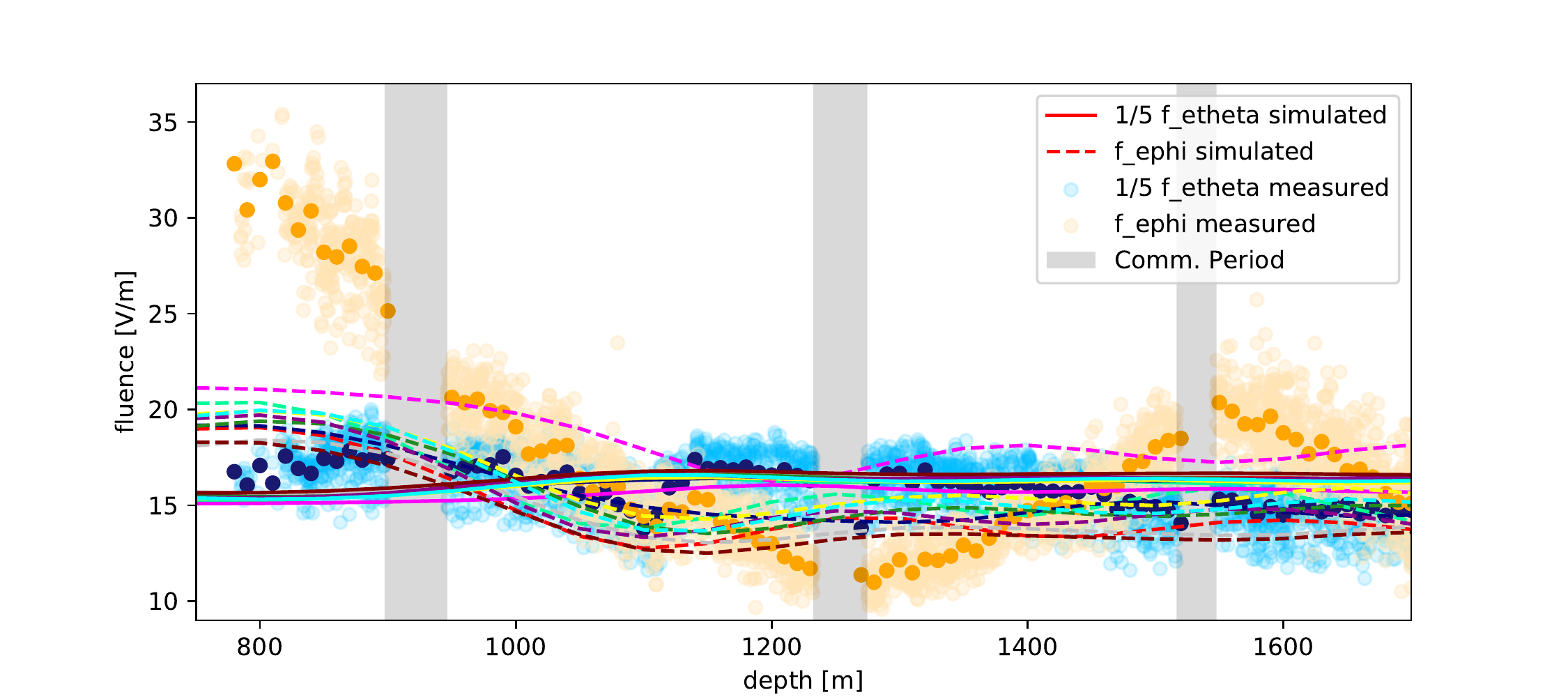}
  \caption{Measured fluence at the ARIANNA station and the expected fluence after propagating the anechoic chamber pulses through the birefringence model. The dominant theta component was downscaled by a factor of 5 for better visibility. The theta and phi components of the ten pulses are represented by solid and dashed lines respectively.}
  \label{fig:fluence}
\end{figure*}

\begin{figure}[tbp]
  \centering
  \includegraphics[width = .48\textwidth]{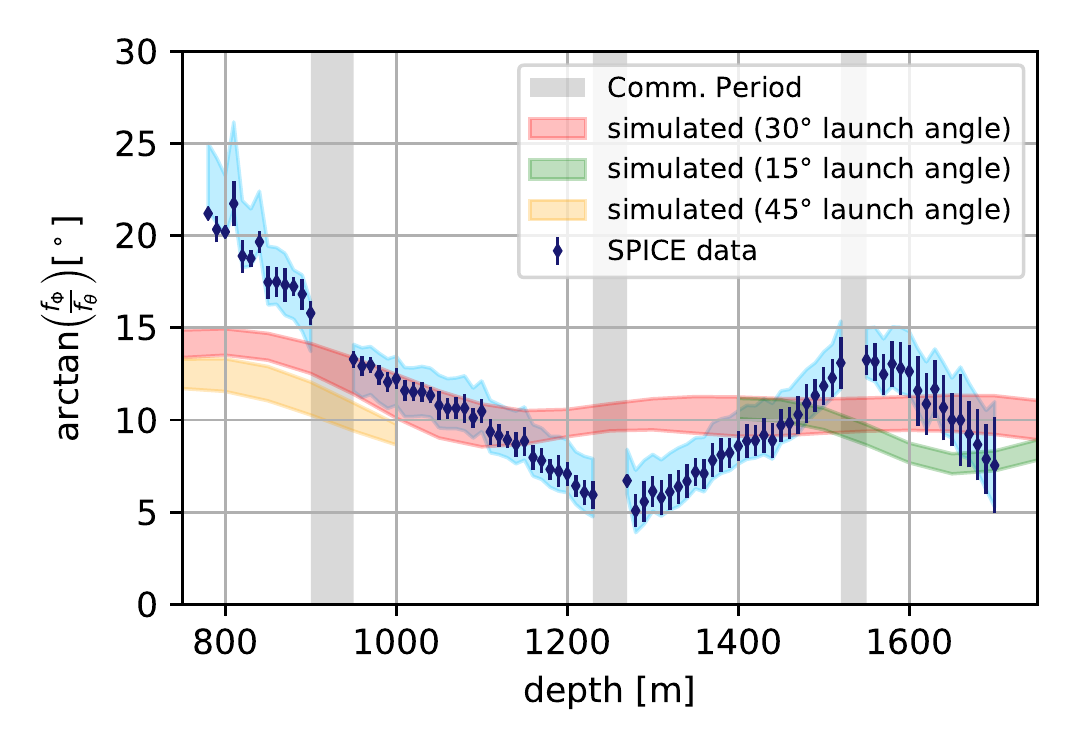}
\caption{Measured polarization at the ARIANNA station and the expected polarization after propagating the anechoic chamber pulses through the birefringence model.The adjacent launch angels of 15 and 45 degrees were included for comparison.}
\label{fig:polarization_angle}
\end{figure}

The ARIANNA collaboration presented the measured fluence of the $\theta$ and $\phi$ polarization components as a function of emitter depth \cite{G_thesis} (which we show in Fig.~\ref{fig:fluence}) and converted it into signal polarization \cite{G_thesis,ARIANNA2020Polarization} which we show in Fig.~\ref{fig:polarization_angle}.
When plotting the measured fluence against the transmitter depth, the data showed depth-dependent variations in the sub-dominant phi component which also translated into a variation of the polarization measurement against depth. In the following, we investigate if the measured variation could originate from birefringence effects. 

We use the pulses measured in the anechoic chamber, propagate them with our birefringence code, and calculate the expected fluence and polarization. 
The total fluence was normalized to the measured average and the results from the 10 pulses at $30^\circ$ launch angle were plotted against the emitter depth. As shown in Fig.~\ref{fig:fluence}, one can observe depth-dependent changes in the fluence but significantly weaker than the variations measured by ARIANNA. The same can be seen in the polarization in Fig.~\ref{fig:polarization_angle}, where the bands describe the statistical uncertainty from the ten different pulses. 

To test the robustness of the model and see if the data could be modeled with the pulses of the adjacent launch angles ($15^\circ$ and $45^\circ$, cf. Fig.~\ref{fig:pol_vs_launch} these pulses were propagated as well and the results are shown in Fig.~\ref{fig:polarization_angle}. We find small differences in the predicted polarization of a few degrees, and for the deepest depths also a different behavior with depth. This indicates that variations of the emitter with different launch angles can have a significant effect on the measurement. As the conditions in the field deviate from the anechoic chamber measurement, it seems plausible that the emitter could be the cause of the observed variations of signal polarization with depth. 

\begin{figure}[tbp]
  \centering
\includegraphics[width = .48\textwidth]{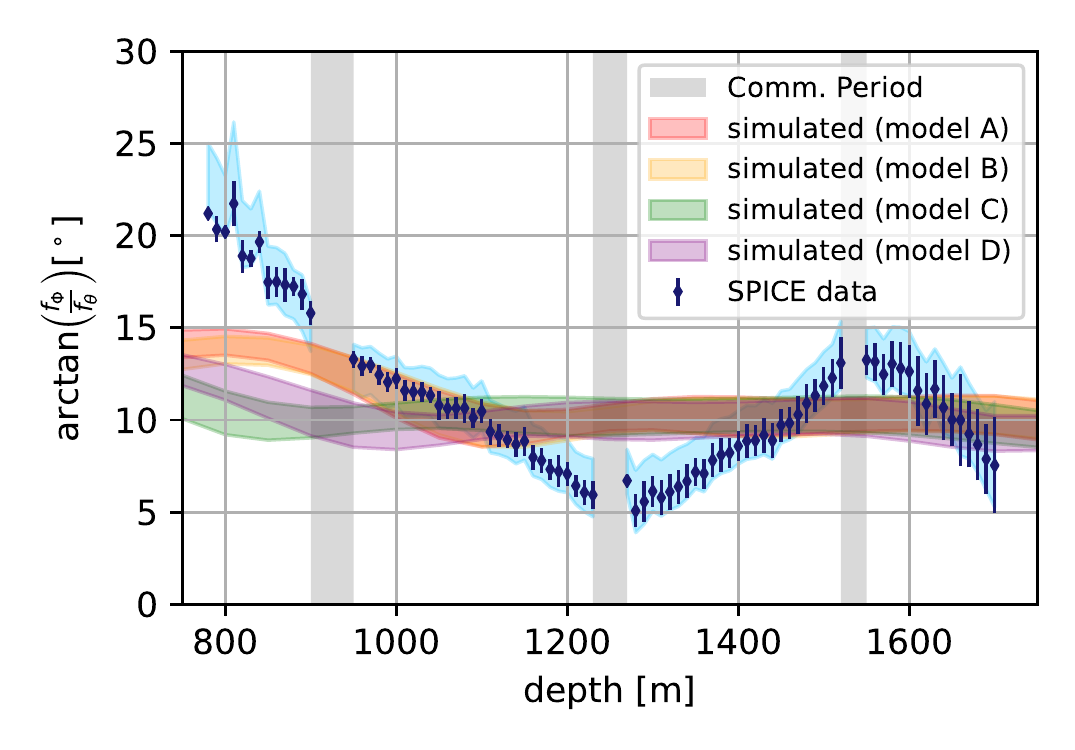}
\caption{Measured polarization at the ARIANNA station and the expected polarization after propagating the anechoic chamber pulses through the birefringence model. The different bands indicate the different ice models tested in the birefringence model~\ref{appndix1}.}
\label{fig:polarization_model}
\end{figure}

We further investigated if we could adapt the parameters of the dielectric tensor to better describe the measured polarization. We repeated the polarization prediction using different interpolations of the dielectric tensor data that are presented in Appendix \ref{appndix1}. The predicted polarization for the different ice models is shown in Fig.~\ref{fig:polarization_model}. None of the adjustments came close to modeling the measured oscillations. Thus, it seems unlikely that the majority of the observed variations in polarization are due to birefringence.

\section{Relevance for neutrino detection}

Birefringence is beneficial for in-ice radio detection but also adds additional challenges to the event reconstruction. It is beneficial because the time delay between the two polarization eigenvectors is linearly proportional to the propagation time. It will allow determining the distance to the neutrino interaction which is needed for reconstructing the neutrino energy. It will work especially well for far-away neutrino interactions where the measurement of the distance through the curvature of the wavefront deteriorates. 
On the other hand, the displacement of pulses reduces the overall amplitude which will reduce the trigger efficiency and therefore the neutrino effective volume of an in-ice neutrino detector. More importantly, due to the change of the polarization eigenvectors during propagation and subsequent interference, the signal polarization can get altered which is a problem because the signal polarization is needed to determine the neutrino direction. However, this is mostly a problem if the time delay is so small that birefringence effects can't be disentangled from the measurement itself, or if systematic uncertainties in the birefringence modeling (i.e. uncertainties in the $\vec{n}(z)$ profile) don't allow to correct for it. 

\begin{figure*}[tbp]
  \centering
  \includegraphics[width = 0.9\textwidth]{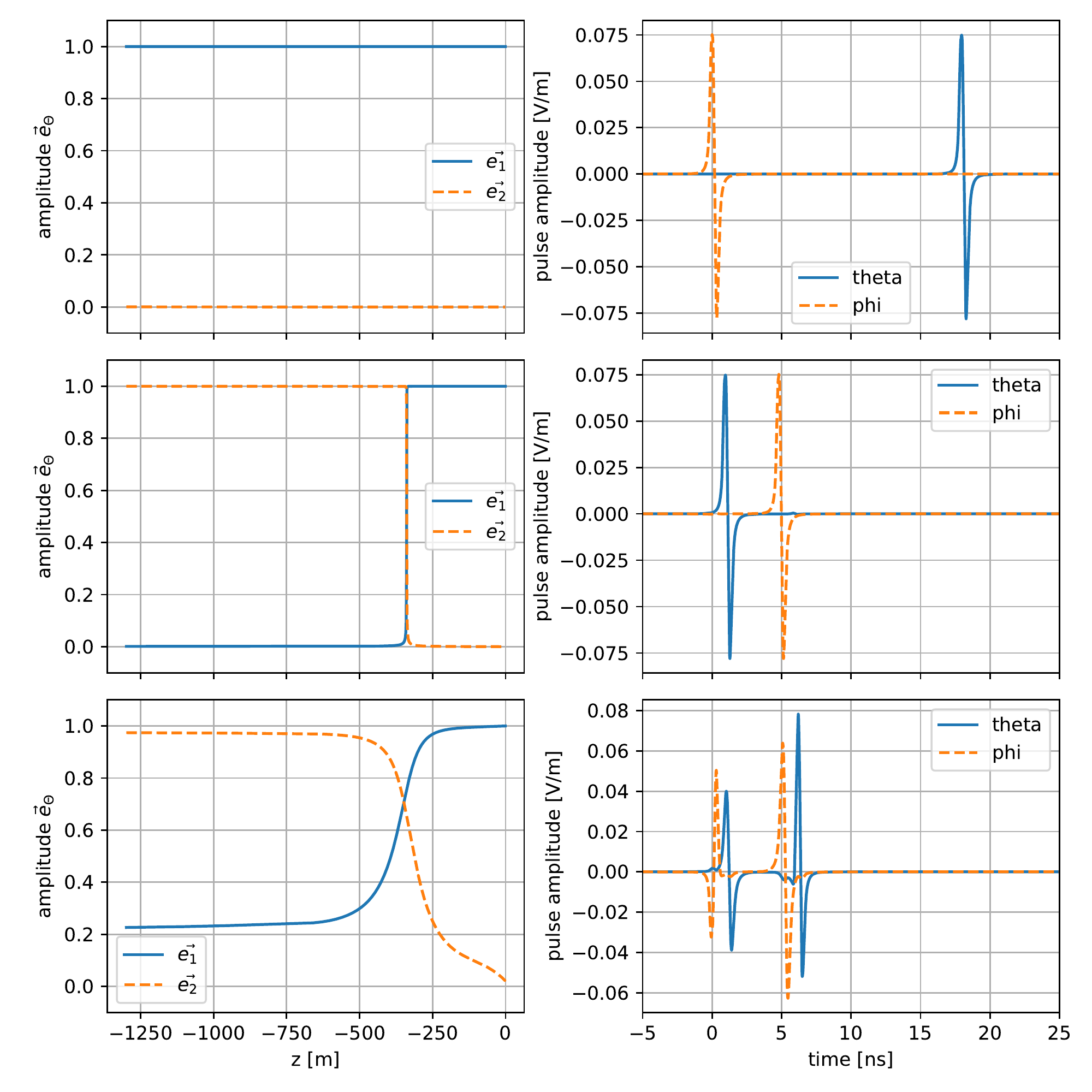}
\caption{Typical pulse shapes generated from NuRadioMC and propagated through the introduced birefringence model. The left side shows the polarization of the the two propagating states against the depth. The right side shows the expected traces after propagating from to the given antenna positions. The neutrinos source for all plots was set to [\SI{0}{m}, \SI{0}{m}, \SI{-1300}{m}]. The top geometry corresponds to an antenna position of [\SI{1}{m}, \SI{1500}{m}, \SI{-1}{m}]. The center geometry corresponds to an antenna position of [\SI{1500}{m}, \SI{1}{m}, \SI{-1}{m}]. The bottom geometry corresponds to an antenna position of [\SI{1487}{m}, \SI{200}{m}, \SI{-1}{m}].}
\label{fig:exp_traces}
\end{figure*}

A systematic study of how birefringence will affect the performance of a radio detector at the South Pole is beyond the scope of this article, but we discuss three typical scenarios in the following that allow estimating the effect on neutrino detection. We propagate an Askaryan pulse from \SI{1300}{m} depth to a shallow antenna at \SI{1}{m} depth that is \SI{1500}{m} away horizontally. We use the predicted electric field for a \SI{e18}{eV} hadronic particle cascade observed at \SI{1}{\degree} away from the Cherenkov cone as the initial pulse and set the signal polarization to have the same amplitude in the theta and phi state. The choice corresponds to a typical geometry expected for neutrino measurements at the South Pole. We study three cases where 1) the propagation is perpendicular to the ice flow, 2) the propagation is parallel to the ice flow, 3) the propagation is at \SI{7.7}{\degree} to the ice flow. The resulting polarization eigenvectors of the two effective index-of-refraction states $N_1$ and $N_2$ as a function of depth and the electric field after propagation at the antenna is shown in Fig.~\ref{fig:exp_traces}. 

For the first case of propagation perpendicular to the ice flow, the polarization eigenvectors align with the theta and phi states. Hence, the result is a clear separation between the radio pulse in the theta and phi components. This geometry is ideal to determine the distance to the neutrino interaction via the birefringence time delay. 
For the second case of a propagation parallel to the ice flow, the two polarization eigenvectors also stay mostly constant but swap over a narrow depth interval. Therefore, no interference and change in polarization takes place and the accumulated time delay is less. This geometry would also allow determining the distance to the neutrino interaction quite easily but with larger uncertainties due to the smaller time delay. 

The vast majority of geometries however lie somewhere in between these 'special' cases and one expects some change of pulse shape and amplitude as well as a substantial time shift between the two states. We show a geometry where this effect is exemplified in Fig.~\ref{fig:exp_traces}. 
Further investigation is needed to quantify how these would affect the sensitivity and reconstruction performance of in-ice neutrino detectors. 

\section{Conclusions}
We calculated the effect of birefringence on in-ice radio propagation from first principles where the only free parameters are the ice properties, i.e., the dielectric tensor at every position in the ice. We combined these results with a numerical propagation code that allows to propagate arbitrary waveforms through the ice. For the first time, this allows making realistic predictions of birefringence effects for arbitrary geometries and arbitrary waveforms. Our code is available open-source through NuRadioMC and can be directly integrated into MC simulations of radio neutrino detectors. 

During propagation, the radio signal splits up into two orthogonal components with slightly different indices of refraction. Because the ice properties and the propagation direction change during propagation, also the effective indices-of-refraction as well as the polarization eigenvectors can change during the propagation. Therefore, in addition to just accumulating a time delay between polarization states, interference takes place which can alter the pulse form and signal polarization. The effect depends strongly on the considered geometry and the initial signal pulse.

We used our birefringence code to make detailed predictions of in-situ measurements of in-ice propagation that are sensitive to birefringence effects conducted by the ARA and ARIANNA collaborations at the South Pole. We base the prediction on measurements of the dielectric tensor from ice fabric measurements, and an anechoic chamber measurement of the signal emitter that was used in the measurement campaign. We found that taking into account the emitted pulse shapes is crucial for interpreting existing in-situ measurements, an effect that was ignored in previous studies of birefringence. After taking into account these effects, we find an agreement between our prediction and the ARA measurement of the time delay between the vertical and horizontal signal components. The measurement of time delays is useful to probe birefringence effects and the agreement we find is an encouraging test of the propagation code but additional measurements for different geometries are needed to better probe the predictive power. 

We also compare our predictions to amplitude and/or polarization measurements where some variation with emitter depth was observed. In previous work that assumed the emission of continuous waves of fixed frequency, it was speculated that the observed amplitude variation stems from birefringence. With our more detailed calculation based on first principles and using a detailed model of the emitted waveforms, we fail to generate amplitudes variations at the level that was observed. We conclude that it is unlikely that birefringence causes this effect. 

Birefringence is beneficial for in-ice radio detection but also adds additional challenges to the event reconstruction. The time delays between polarization states give access to the distance to the neutrino interaction which is needed to estimate the neutrino energy. On the other hand, a change in signal polarization will complicate the reconstruction of the neutrino direction. 
A systematic study of the effect of birefringence on in-ice radio detection is beyond the scope of this article but we discussed three different geometries that envelop typical cases. For propagation direction along and perpendicular to the ice flow, we observe only an accumulation of a time delay without any change to the pulse shapes or amplitude. For geometries in between, the polarization eigenstates show a smooth transition over longer propagation length which leads to interference effects. A typical result is a double pulse in both polarization states. 
In future work, we will systematically study the effect of birefringence on in-ice radio detection of high-energy neutrinos. The detailed predictions that can be made using this work can be used to develop reconstruction algorithms that exploit birefringence effect for energy and direction reconstruction, e.g., through the use of deep neural networks. 

Additional measurement campaigns at the South Pole sensitive to birefringence effects will be useful to solidify the predictive power of the birefringence calculation presented here. Work to integrate the propagation code into RadioPropa is ongoing which will allow propagation in media with arbitrary $\vec{n}(x,y,z)$. In addition, the remaining approximation of modeling the propagation using ray optics can be checked via FDTD simulations but due to their extreme computational costs, it would only be feasible to test a few selected geometries.

\section{Acknowledgments}
We thank Dave Besson, Anna Nelles, and Bob Oeyen for feedback on the manuscript. We thank all members of the inicemc working group for the feedback on this work. We thank the developers of the NuRadioMC code for their help in integrating the birefringence model into NuRadioMC.

\bibliographystyle{JHEP}
\bibliography{bib}

\appendix

\section{Different Ice Models}
\label{appndix1}

Different experiments report on different density profiles such that for the ARIANNA measurements the parameters of Eq.~\eqref{eq:density} were $n_{ice}= 1.78$, $\Delta n = 0.426$, $z_0= \SI{77}{\meter}$ \cite{southpole_2015} and for the ARA measurements the parameters were $n_{ice}= 1.78$, $\Delta n = 0.454$, $z_0=  \SI{49.504}{\meter}$ \cite{ARA_2022}. For the analysis of the data by the two collaborations, the respective models were used. 
 
\begin{figure}[tbp]
  \centering
  \includegraphics[width = 0.5\textwidth]{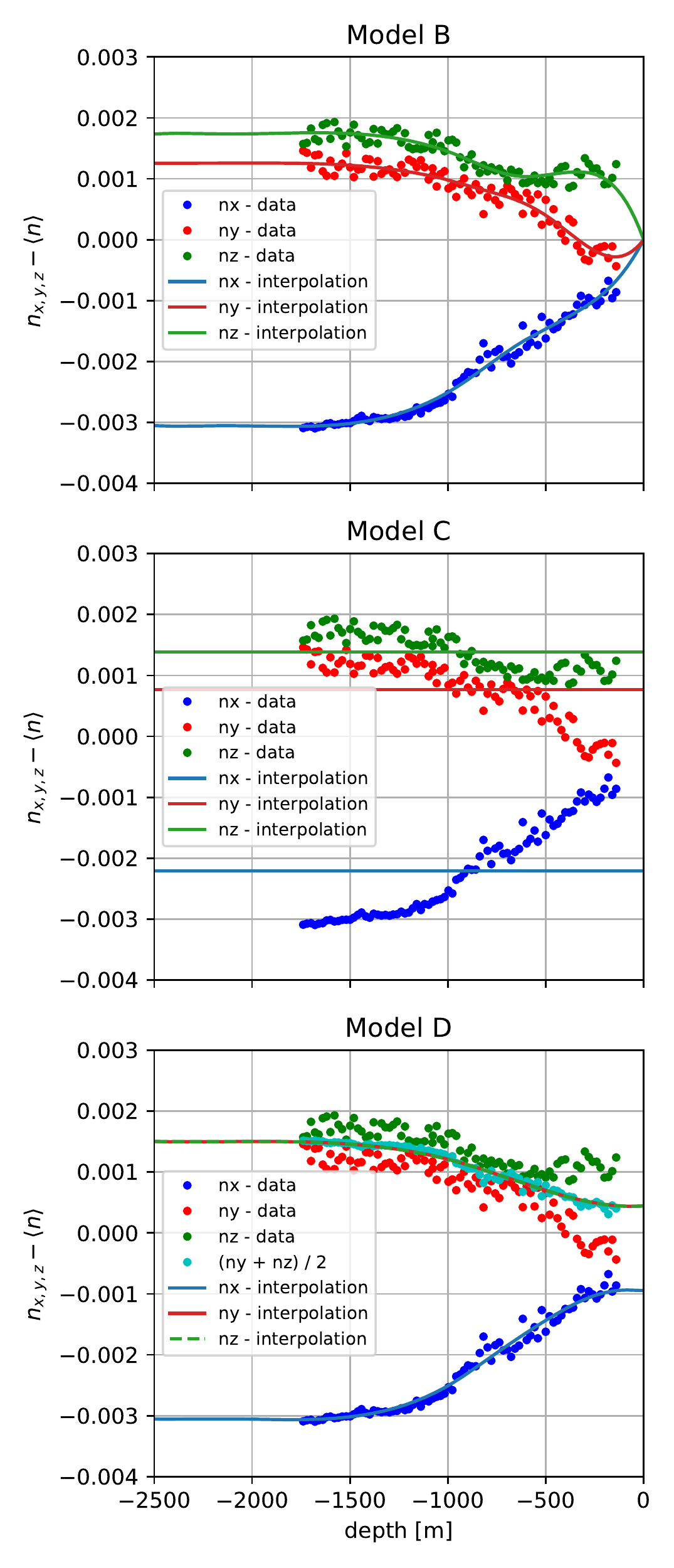}
  \caption{Refractive index as a function of depth. Measured birefringence data (scatter) \cite{voigt} compared to the average index-of-refraction value at this depth of $\langle n(z) \rangle$.. The shown spline interpolation of the data (solid lines) was used in the analysis to create figure \ref{fig:polarization_model}. Model B assumes a converging index of refraction at shallow depths. Model C is a constant average over all depths. Model D assumes $n_y$ and $n_z$ to be the same vale at the average of the two.}
  \label{fig:birefringence_model}
\end{figure}

Figure \ref{fig:birefringence_model} shows the different interpolations used to see how big the changes affect the polarization measurement of ARIANNA in \ref{fig:polarization_model}. 

\section{Effective Refractive Indices}
\label{appndix2}

The two effective refractive indices are found by calculating the roots of Eq.~\eqref{eq:n}. As it is a six order polynomial, numerical methods were initially used to find the roots. However, this is computationally expensive and it is possible to find analytical solutions for the roots. When using the normalization of the direction vector, the 6th-order term of Eq.~\eqref{eq:n} vanishes and when substituting $r$ for $n^2$ it reduces Eq.~\eqref{eq:n} to a simple quadratic equation with two roots $R_{1,2}$. Reversing the substitution and ignoring the nonphysical negative solutions returns the two effective indices $N_{1,2}$:

\begin{equation} 
\begin{split}
\label{eq:roots_r}
R_{1, 2} =\left.  \big(-2{n_x}^2 {n_y}^2 {n_z}^2\big) \middle/ \right.\\
\bigg({n_y}^2 {n_z}^2 \big(-1+{s_x}^2\big)+{n_x}^2 \big({n_z}^2 \big(-1+{s_y}^2\big)\\
+{n_y}^2 \big(-1+{s_z}^2\big)\big) \\
\pm \big( 4{n_x}^2 {n_y}^2 {n_z}^2 \left({n_z}^2 \big(-1+{s_x}^2+{s_y}^2\right) \\
+{n_y}^2 \big(-1+{s_x}^2+{s_z}^2\big)+{n_x}^2 \big(-1+{s_y}^2+{s_z}^2\big)\big) \\
+\big({n_y}^2 {n_z}^2 \big(-1+{s_x}^2\big)+{n_x}^2 \big({n_z}^2 \big(-1+{s_y}^2\big)\\
+{n_y}^2 \big(-1+{s_z}^2\big)\big)\big)^2 \big)^{\frac{1}{2}}\bigg) 
\end{split}
\end{equation}

\begin{equation} 
\label{eq:roots_n}
N_{1,2} =  \sqrt{R_{1,2}}
\end{equation}

\end{document}